\documentclass[english,runningheads,a4paper]{llncs}

\usepackage[utf8]{inputenc}
\usepackage[OT1]{fontenc}
\usepackage{babel}
\usepackage{csquotes}
\usepackage[final]{graphicx}

\usepackage{ifdraft}
\usepackage{enumitem}
\usepackage{cite}
\usepackage{rotating}

\usepackage{xspace}

\usepackage{amsmath}
\usepackage{amssymb}

\usepackage{booktabs}
\usepackage{multirow}
\usepackage{tabularx}
\newcolumntype{W}{>{\raggedleft\arraybackslash}X}
\newcolumntype{T}{>{\centering\arraybackslash}X}

\usepackage{complexity}

\newlength{\primadiequazioni}
\newlength{\dopodiequazioni}
\newcommand{\compattaspazioequazioni}{%
  \setlength{\primadiequazioni}{\abovedisplayskip}%
  \setlength{\dopodiequazioni}{\belowdisplayskip}%
  \setlength{\abovedisplayskip}{-1ex}%
  \setlength{\belowdisplayskip}{0pt}%
}
\newcommand{\ripristinaspazioequazioni}{%
  \setlength{\abovedisplayskip}{\primadiequazioni}%
  \setlength{\belowdisplayskip}{\dopodiequazioni}%
}

\usepackage[centerlast]{subfigure}

\usepackage[final]{hyperref}

\newcommand{\inputdata}[1]{\noindent \emph{Input: }#1\\*}
\newcommand{\outputdata}[1]{\noindent \emph{Output: }#1}
\newcommand{\otoprule}{\midrule[\heavyrulewidth]}

\newcommand{\ie}{i.e.,~}

\renewcommand{\emptyset}{\ensuremath{\varnothing}}
\newcommand{\Z}{\ensuremath{\mathbb{Z}}}

\newcommand{\probl}[1]{\textsc{#1}\xspace}

\newcommand{\MRHC}{\textsc{MRHC}\xspace}
\newcommand{\HCRE}{\textsc{HCRE}\xspace}
\renewcommand{\SAT}{\textsc{SAT}\xspace}
\newcommand{\HCstar}[2]{\textsc{\ensuremath{(#1,#2)}-HC}\xspace}

\newcommand{\reHCstar}{\HCstar{r}{e}}

\newcommand{\lcom}[1]{(\textit{#1})\xspace}
\newcommand{\li}{\lcom{i}}
\newcommand{\lii}{\lcom{ii}}
\newcommand{\liii}{\lcom{iii}}

\newcommand{\progr}{\textsc{reHCstar}\xspace}
\newcommand{\eps}{\ensuremath{\varepsilon}}

\newcommand{\testotitolo}{Haplotype Inference on Pedigrees with
  Recombinations, Errors, and Missing Genotypes via SAT solvers}
\newcommand{\testoautore}{%
  Yuri Pirola, Gianluca Della Vedova,
  Stefano Biffani,
  Alessandra Stella, and
  Paola Bonizzoni}
\newcommand{\testointest}{\reHCstar via SAT}

\hypersetup{%
  colorlinks=false,
  pdftitle    = {\testotitolo},
  pdfsubject = {},
  pdfkeywords = {pedigree haplotyping haplotype inference recombinations
  mutations errors missing genotypes},
  pdfauthor   = {\testoautore},
  pdfcreator = {Yuri Pirola},
  pdfproducer = {},
  pdfborder = 0 0 0
}

\numberwithin{equation}{section}

\begin{document}

\title{\testotitolo}
\titlerunning{\testointest}
\author{Yuri Pirola\inst{1,2} \and
  Gianluca Della Vedova\inst{3} \and
  Stefano Biffani\inst{2} \and
  Alessandra Stella\inst{2,4} \and
  Paola Bonizzoni\inst{1}
}
\authorrunning{Pirola \textit{et al.}}
\institute{%
DISCo, Univ.~degli Studi di Milano--Bicocca, Milan, Italy,
\email{\{pirola,bonizzoni\}@disco.unimib.it}
\and
CeRSA, Parco Tecnologico Padano, Lodi, Italy,
\email{\{stefano.biffani,alessandra.stella\}@tecnoparco.org}
\and
Dip. Statistica, Univ.~degli Studi di Milano--Bicocca, Milan, Italy,
\email{gianluca.dellavedova@unimib.it}
\and
IBBA, Consiglio Nazionale delle Ricerche, Lodi, Italy.
}

\maketitle

\begin{abstract}
The \probl{Minimum-Recombinant Haplotype Configuration} problem (\MRHC)
has been highly successful in providing a sound combinatorial
formulation for the important problem of genotype phasing on pedigrees.
Despite several algorithmic advances and refinements that led to some
efficient algorithms, its applicability to real datasets has been
limited by the absence of some important characteristics of these data
in its formulation, such as mutations, genotyping errors, and missing data.

In this work, we propose the \probl{Haplotype Configuration with
Recombinations and Errors} problem (\HCRE), which generalizes the
original \MRHC formulation by incorporating the two most common
characteristics of real data: errors and missing genotypes (including
untyped individuals).
Although \HCRE is computationally hard, we
propose an exact algorithm for the problem based on a reduction to the
well-known Satisfiability problem.
Our reduction exploits recent progresses in the constraint programming
literature and, combined with the use of state-of-the-art SAT solvers,
provides a practical solution for the \HCRE problem.
Biological soundness of the phasing model and effectiveness (on both
accuracy and performance) of the algorithm are experimentally
demonstrated under several simulated scenarios and on a real dairy
cattle population.
\end{abstract}

\section{Introduction}

After the first draft of the human genome was published in 2000, a huge
research effort has been devoted to the discovery of genetic
differences among same-species individuals and to the
characterization of their impact to the expression of different phenotypic
traits such as disease susceptibility or drug resistance.
Moreover, the completion of several livestock genomes and the
recent introduction of genomic data into breeding programs widened the
interest of these studies to other species.
Most of these efforts are driven by the \emph{International HapMap
Project}~\cite{haplomapII}, which discovered, investigated and
characterized millions of genomic positions (called \emph{loci} or
\emph{sites}) where different individuals carry different genetic
subsequences (called \emph{alleles}).
In practice, unordered pairs of alleles coming from both parents of each
individual studied are
routinely collected, since determining the parental source of each
allele is too time-consuming and expensive to be performed on large
studies~\cite{BonizzoniVDL03}.
The pairs of alleles located at a given set of loci of an individual are
called the
(multi-locus) \emph{genotype} of the individual, while the sequence of
alleles that were inherited from a single parent is called a
\emph{haplotype}.
The advance of high-throughput and high-density 
genotyping
technologies, combined with a consistent reduction of genotyping costs,
has led to a great abundance of genotypic data.
Such genotypes (also called SNP genotypes) are generally biallelic (\ie
at each locus only two distinct alleles are observed in the population)
and they will be the focus of this work.
A number of association studies based on SNP genotypes have been carried
out but, since haplotypes substantially increase the power of genetic
variation studies~\cite{Tregouet2009}, accurate and efficient
computational prediction of haplotypes from genotypes is highly desirable.
Mendelian inheritance laws, which govern the transmission of genetic
material from parents to children, have been effectively used to
improve the accuracy of haplotyping methods.
The general problem that we have just described is called in the literature
Haplotype Inference (HI) on pedigrees, and it  asks
for a haplotype configuration consistent with a given genotyped
pedigree.
While the problem has been successfully tackled by classic
statistical methods (such as Lander-Green~\cite{LanderGreen} and
Elson-Stewart~\cite{ElsonStewart}), the increasing density and length of SNP
genotypes pose new hurdles. In fact those methods are not designed to scale
well on large datasets and they do not take directly into account the
presence of Linkage Disequilibrium among loci in the founder population.

Combinatorial formulations have been proposed to overcome such
limitations.
Since there can exist an exponential number of consistent
haplotype configurations, we need
an additional criterion for choosing the ``right'' haplotype
configuration among those compatible with the input data.
The genetic linkage between neighbouring loci has inspired
a parsimonious formulation of the HI problem that looks for a configuration
minimizing the number of recombination events in the
resulting haplotyped pedigree.
This formulation, called \probl{Minimum-Recombinant Haplotype
Configuration} (\MRHC)~\cite{Qian2002,Li03efficientinference}, has been
shown to be a successful approach.
The aim of this formulation is the computation of a haplotype
configuration which is consistent with an input genotyped pedigree and
induces the minimum number of recombinations.
The formulation naturally arises since recombinations are the most
common source of genetic variation.
However, the HI problem where mutations were allowed instead of
recombinations has been solved by an ILP-based
algorithm~\cite{mephase}.
An efficient heuristic when  both recombinations and
mutations are allowed has been presented in~\cite{mchc-tcbb11}.

Despite several remarkable advances in genotyping technologies,
genotypes are usually affected by a small percentage of errors and missing
data which, even at very small rates ($<0.5\%$), could heavily affect the outcome of subsequent
analyses.
Moreover, missing genotypes could represent a significant portion
($5\%$ or more) of the dataset due to uncertainty in genotype call procedures
or unavailability of the DNA sample of some individuals of the pedigree.
These aspects shared by almost all actual datasets were not
considered in the original \MRHC formulation, limiting its applicability
and practical relevance.

In this paper, we propose the \probl{Haplotype Configuration with
Recombinations and Errors} (\HCRE) problem, which generalizes the \MRHC
formulation by allowing genotyping errors and missing data in addition
to recombination events.
Polynomial-time exact algorithms for \HCRE are unlikely to exist since
the problem is \APX-hard even on simple
instances.
The main contribution of this paper is a practical exact algorithm for
\HCRE, based on a reduction from \HCRE to the well-known Satisfiability
problem (SAT), for which extremely efficient solvers are known.
Our reduction exploits some characteristics of one of the best
performing SAT solvers, CryptoMiniSat~\cite{CryptoMiniSat}, in order to
compute a solution for most of the practical instances with modest
computing resources.
An extensive experimental evaluation of our algorithm under several simulated
scenarios and on a real dairy cattle population demonstrates its
accuracy and performance.

\section{The Computational Problem}
\label{sec:comp-probl}

In this section we define the basic notions
that will be studied in the rest of the
work.
A \emph{pedigree graph} is an oriented acyclic graph $P=(V,E)$ such that
\li vertices correspond to individuals and are partitioned into
\emph{male} and \emph{female} vertices (\ie $V=M \cup F$, $M \cap
F=\emptyset$),
\lii each vertex has indegree 0 or 2, and
\liii if a vertex has indegree 2, then one edge must come from a male
node and the other from a female node.
If a vertex $v$ has indegree $0$, then $v$ is called
\emph{founder}.
For each edge $(p, c) \in E$, we say that $p$ is a parent of $c$ and $c$
is a child of $p$.
More precisely,  $p$ is the father (mother, resp.) of $c$ if
$p$ is male (female, resp.).

The (possibly incomplete) \emph{genotype} of an individual $i$ is a
$n$-long vector $g_i$ over the set $\{ 0, 1, 2, * \}$ where we follow the
convention of encoding the unordered pair of alleles $\{0,0\}$ as $0$,
$\{1,1\}$ as $1$, and $\{0,1\}$ as $2$, while $*$ represents a missing
(or ``not called'') genotype.
A genotype is \emph{complete} if it does not contain the
$*$ element, otherwise is \emph{incomplete}.
An individual $c$ is said to be \emph{heterozygous} in a given locus $i$
if $g_c[i] = 2$, and \emph{homozygous} if $g_c[i] \in \{0,1\}$.
A \emph{haplotype configuration} $H$ is an assignment of a pair of
haplotypes $(h^0_i, h^1_i)$ to each individual $i$ and a pair of source
vectors $(s_{f,i}, s_{m,i})$ to each non-founder individual $i$ of the
pedigree (where $f$ and $m$ are the parents of $i$).
Both a \emph{haplotype} and an \emph{source vector} are binary $n$-long
vectors. In a haplotype $0$ and $1$ are the two (major and minor)
alleles.
Informally, the source vector $s_{p,i}$ associated with a haplotype
$h^p_i$, where $p$ is a parent of $i$, indicates if the allele
$h^p_i[l]$ has been inherited from the paternal ($s_{p,i}[l]=0$) or
maternal ($s_{p,i}[l]=1$) haplotype of $p$.
More formally, let $h_i=(h^0_i, h^1_i)$ be the haplotypes of a
non-founder individual $i$.
Then $h_i$ is consistent with the Mendelian laws of inheritance for a
locus $l$ if
\lcom{a} $h^0_i[l] = h^{s_{f,i}[l]}_f[l]$ where $f$ is the father of $i$, and
\lcom{b} $h^1_i[l] = h^{s_{m,i}[l]}_m[l]$ where $m$ is the mother of $i$.
Let $h_i=(h^0_i, h^1_i)$ be the haplotypes of an individual $i$ with
genotype $g_i$.
Then $h_i$ is consistent with $g_i$ in a locus $l$ if
$h^0_i[l] = h^1_i[l] =g_i[l]$ if $l$ is homozygous, and $\{ h^0_i[l],
h^1_i[l]\} = \{0,1\}$ if $l$ is heterozygous.
A haplotype $h^p_i$ of individual $i$ inherited from its parent $p$
contains a recombination at locus $ l > 0 $ if $s_{p,i}[l-1] \neq
s_{p,i}[l]$.

In this work, we study the \reHCstar problem, which formalizes the HI
problem on pedigrees allowing the presence of recombinations
and genotype inconsistencies.
If $r$ or $e$ are $*$, then the corresponding values are intended as
\emph{unbounded}.
The \reHCstar problem generalizes  the
\probl{Minimum-Recombinant Haplotype Configuration}
(\MRHC) problem~\cite{Li03efficientinference}, where only recombinations
are allowed and all the genotypes are assumed to be complete and
correctly called.

\begin{problem}\probl{$(r,e)$-Haplotype Configuration
    problem}(\reHCstar).\\*
\inputdata{A genotyped pedigree $P$ with possibly incomplete genotypes
  (\ie $g_i[l] = *$ for some individual $i$ and locus $l$).}
\outputdata{A haplotype configuration $H$ for the genotyped pedigree $P$
  such that:
  \begin{enumerate}[topsep=0pt, partopsep=0pt]
  \item $H$ is consistent with the Mendelian laws of inheritance for
    each individual $i$ and locus $l$;
  \item $H$ is consistent with the observed genotypes in all but at most
    $e$ cases;
  \item $H$ contains at most $r$ recombinations.
  \end{enumerate}
}
\end{problem}

Notice that, while we motivated our problem with a parsimonious
principle,
we preferred not to formulate \reHCstar as an optimization problem.
In fact, any cost function that melds the number of recombinations and
errors (\ie by applying some weights to them) would potentially
introduce some bias in the computed solution.
Instead, the two parameters $r$ and $e$ directly map to two important
characteristics of the dataset, the genetic distance among the markers
and the quality of the collected genotypes, for which the researcher could
provide good estimates.
As an example, the recombination rate between adjacent markers in
high-density panels is about $10^{-6}$ while the genotyping
error rate (after quality check filters) is generally less than $0.5\%$.

\section{Reducing \reHCstar to SAT}
\label{sec:rehctosat}

A main technical device of this work consists of a reduction from
\reHCstar to \SAT.
In this work, an instance of SAT is a set of ``extended clauses'', where
each extended clause is either the disjunction or the exclusive-OR of
literals (\ie variables or their negation).
The instance is satisfiable if and only if all the extended clauses are
satisfiable.
The main reason for choosing such a generalization is that our reduction
is slightly simplified by its use since XORs of literals are additions
over the $\Z_2$ field.
Moreover the SAT solver that we selected,
CryptoMiniSat~\cite{CryptoMiniSat}, is  designed to solve
those instances.
For simplicity, we slightly abuse  the language by indicating also the
exclusive-OR of literals with the term ``clause''.
In the following we denote with $\oplus$  the exclusive-OR,
with $=$ the equivalence,
with $\wedge$ the conjunction,
with $\vee$ the disjunction,
and with $\neg$ the negation.

In order to gently guide the reader, first we present the reduction for
the case of unbounded number of recombinations and errors.
Then, we will deal with the general problem \reHCstar.
In our reduction we use some Boolean variables:
\begin{itemize}
\item $p_i[l]$, $m_i[l]$, alleles of the paternal and maternal
  haplotypes of individual $i$ at locus $l$ (\ie $h^0_i[l]$ and
  $h^1_i[l]$) where \emph{false} is equal to $0$ and
  \emph{true} is equal to $1$;
\item $s_{p,i}[l]$, which is \emph{true} if the source vector of
  individual $i$ (w.r.t.~parent $p$) at locus $l$ is equal to $1$, and
  \emph{false} if it is equal to $0$;
\item $r_{p,i}[l]$, which is \emph{true} if a recombination has occurred
  between individual $i$ and its parent $p$ at locus $l$ (\ie if
  $s_{p,i}[l-1] \ne s_{p,i}[l]$);
\item $e_i[l]$, which is \emph{true} iff the haplotype configuration is
  not consistent with the observed genotypes for individual $i$ at locus
  $l$.
\end{itemize}

To better describe our model and the associated SAT instance, we will put on
the left the constraint we are imposing and on the right the corresponding
clauses, preceded by the condition under which the constraint must hold.
A \HCstar{*}{*} instance can be encoded by the logic formula consisting
of the conjunction of the following clauses which encodes three kinds of
constraints: constraints for Mendelian laws consistency, constraints for
genotype consistency, and constraints for recombination representation.
The first class of constraints ensures the consistency with the
Mendelian laws of inheritance between a non-founder individual $i$ and
its parent $p$.
In other words, each allele of the haplotype of $i$ inherited from $p$
must be equal to one allele of one of the haplotypes of $p$ depending on
the variable $s_{p,i}$.
We abstract from the gender of $p$ and
use  $c_{p,i}[l]$ to indicate the variable $p_i[l]$ if $p$ is
father of $i$ and  $m_i[l]$ if $p$ is mother of $i$.\\[2pt]
\compattaspazioequazioni
\begin{tabularx}{\linewidth}{|T|W|}
\hline
\multicolumn{2}{|l|}{For each individual $i$, each parent $p$ of $i$, and each locus $l$:}\\
\hline
\begin{multline*}
\big( \left( p_p[l] \wedge \neg s_{p,i}[l]\right) \oplus \\
\oplus \left( m_p[l] \wedge  s_{p,i}[l] \right)\big) =
c_{p,i}[l]
\end{multline*}
\vfill
&
\begin{equation}\label{eq:hcstarstar:parent}
\begin{cases}
s_{p,i}[l]\vee      p_p[l]\vee \neg c_{p,i}[l]\\
s_{p,i}[l]\vee \neg p_p[l]\vee      c_{p,i}[l]\\
\neg s_{p,i}[l]\vee      m_p[l]\vee \neg c_{p,i}[l]\\
\neg s_{p,i}[l]\vee \neg m_p[l]\vee      c_{p,i}[l]\\
\neg p_p[l]\vee \neg m_p[l]\vee     c_{p,i}[l]\\
p_p[l]\vee     m_p[l]\vee \neg c_{p,i}[l]\\
    \end{cases}
\end{equation}
\\
\hline
\end{tabularx}
\ripristinaspazioequazioni

Notice that the last two clauses of (\ref{eq:hcstarstar:parent}) are implied
by the other clauses.
However, their explicit inclusion in our formulation improves the
propagating behaviour and the overall efficiency of the SAT
solver~\cite{SATencodings}.

The second class of constraints ensures that the computed haplotypes are
either consistent with the observed genotypes or that variable $e_i[l]$
is \emph{true}.
This leads to three different equations depending on the observed
genotype (clearly no constraint is needed for an individual $i$ and
locus $l$ where the genotype $g_i[l]$ is missing).

\noindent\compattaspazioequazioni
\begin{tabularx}{\linewidth}{|T|W|}
\hline
\multicolumn{2}{|l|}{For each individual $i$ and locus $l$ such that $g_i[l]=0$ :}\\
\hline
\begin{equation*}
e_i[l] \neq ( p_i[l] = m_i[l] = 0 )
\end{equation*}
&
\begin{equation}\label{eq:hcstarstar:base0}
\begin{cases}
\neg e_i[l] \vee      p_i[l] \vee m_i[l]\\
e_i[l] \vee \neg p_i[l]\\
e_i[l] \vee \neg m_i[l]\\
\end{cases}
\end{equation}
\\
\hline
\end{tabularx}

\noindent
\begin{tabularx}{\linewidth}{|T|W|}
\hline
\multicolumn{2}{|l|}{For each individual $i$ and locus $l$ such that $g_i[l]=1$ :}\\
\hline
\begin{equation*}
e_i[l] \neq ( p_i[l] = m_i[l] = 1 )
\end{equation*}
&
\begin{equation}\label{eq:hcstarstar:base1}
\begin{cases}
\neg e_i[l] \vee \neg p_i[l] \vee \neg m_i[l]\\
e_i[l] \vee      p_i[l]\\
e_i[l] \vee      m_i[l]\\\end{cases}
\end{equation}
\\
\hline
\multicolumn{2}{|l|}{For each individual $i$ and locus $l$ such that $g_i[l]=2$ :}\\
\hline
\begin{equation*}
e_i[l] = (p_i[l] = m_i[l])
\end{equation*}
&
\begin{equation}\label{eq:hcstarstar:base2}
e_i[l] \oplus p_i[l] \oplus m_i[l]
\end{equation}\\
\hline
\end{tabularx}
\ripristinaspazioequazioni

The last class of constraints ensures that $r_{p,i}[l]$ is \emph{true}
when there is a recombination  between individual
$i$ and its parent $p$ at locus $l$ (\ie $s_{p,i}[l-1] \ne
s_{p,i}[l]$).\\
\compattaspazioequazioni
\begin{tabularx}{\linewidth}{|T|W|}
\hline
\multicolumn{2}{|l|}{For each individual $i$, each parent $p$ of $i$, and each locus $l>1$:}\\
\hline
\begin{equation*}
r_{p,i}[l] = (s_{p,i}[l-1] \ne s_{p,i}[l])
\end{equation*}
&
\begin{equation}\label{eq:hcstarstar:recomb}
\neg r_{p,i}[l] \oplus s_{p,i}[l-1] \oplus s_{p,i}[l]
\end{equation}
\\
\hline
\end{tabularx}
\ripristinaspazioequazioni

The conjunction of all previous clauses is
an instance of \HCstar{*}{*}.
Notice that
the total number of variables and clauses is
$O(nm)$, where $n$ is the pedigree size and $m$ the length of the
genotype.

To bound the total number of errors and recombination, we simply have to add
two cardinality constraints:
\begin{equation}\label{eq:hczeroe:base}
  \sum_{\substack{%
    \text{individual }i\\
    \text{locus }l
  }} \!\!\!\!\!\! e_i[l] \le e
  \qquad\qquad
  \sum_{\substack{%
    \text{individual }i\\
    \text{parent }p\text{ of }i\\
    \text{locus }l
  }} \!\!\!\!\!\! r_{p,i}[l] \le r
\end{equation}

Converting a cardinality constraint $\sum_{1 \le i \le n} v_i \le k$ in
a compact Boolean formula is not straightforward since a na\"ive
approach would consider all the subsets of $k$ elements and, thus, would
produce an exponential number of clauses, ruling out even moderate
instances.
This problem of devising an ``efficient'' encoding of a cardinality
constraint into a CNF formula has been investigated in the constraint
programming literature and a promising approach has been recently
proposed.
This technique is based on the construction of \emph{cardinality
networks}~\cite{Asin11} and essentially sorts the set $\{ x_1, \ldots,
x_n\}$ of variables composing the constraint obtaining a
permutation $\left< y_1, \ldots, y_n\right>$ such that whenever
$y_i$ is false, then all variables $y_j$ for $j>i$ are also false.
Consequently, bounding the number of recombinations and errors consists of
forcing the variables $y_{r+1}$ or $y_{e+1}$ to be \emph{false}.
Two advantages of the cardinality network approach are that only
$O(n \log^2 k)$ additional clauses are required and that arc-consistency
under unit propagation is preserved, which improves the performances of modern
SAT solvers.
We refer the  reader
to~\cite{Asin11} for the detailed description of the required clauses..

\section{Experimental Results}
\label{sec:experim}

An implementation of our algorithm, called \progr,
is available at
\url{http://www.algolab.eu/reHCstar}.
We used our implementation to investigate
feasibility, accuracy and performance of our approach under several scenarios.
First, we analyzed the effect of changing the main
parameters (such as pedigree size, missing rate, etc.) on the accuracy
and the performance of our algorithm.
Second, we compared \progr with MePhase~\cite{mephase}, which is a
combinatorial approach to a slightly different formulation of the
Haplotype Inference problem on pedigrees.
Third, we present the evaluation of \progr on a complex
pedigree of a real bovine population.
We  tested on such pedigree  the soundness of the
\reHCstar formulation and the ability of \progr to handle these
instances.

\subsubsection*{Evaluation on Random Instances.}
\label{sec:experim:rnd}

We evaluated how the main problem parameters --
pedigree size ($n$),
genotype length ($m$),
recombination probability ($\theta$),
error probability ($\eps$),
and missing probability ($\mu$) --
affect
the accuracy and the performance of \progr.
For each choice of the parameters, we generated 10 different random pedigree
graphs. Then we generated 10 random haplotype
configurations for each pedigree as follows.
Two haplotypes have been randomly generated for each founder of the
pedigree.
Haplotypes of non-founder individuals have been uniformly sampled from
those of their parents and a recombination has been applied at each
locus with probability $\theta$.
Genotypes of the individuals are then computed from their haplotypes.
Finally, each locus of each individual genotype has been replaced with a
different pair of alleles with probability $\eps$
and  ``masked'' with probability $\mu$ (to simulate missing genotypes).
Accuracy of the results computed by \progr on each instance has been
evaluated with respect to the original haplotype configuration according
to \emph{genotype imputation error rate} (the fraction of the
missing genotypes that have been incorrectly imputed) and
\emph{haplotyping error rate} (the fraction of alleles that have been  incorrectly predicted).
The latter ratio  has been computed for the set of all loci and for the set of
non-missing genotypes.
Performance has been evaluated considering the \emph{average} and the
\emph{maximum running time} of \progr on a standard
workstation with a 2.8GHz CPU.
We have chosen a base set of reasonable values for the parameters:
$n=50$,
$m=50$,
 $\theta=0.005$,
 $\eps=0.005$,
and $\mu=0.05$.
We performed five series of tests; in each of these series only one parameter
is allowed to assume different values, while the other four parameters
are fixed to the base value.
Moreover we fixed the values of $r$ and
$e$ on each instance as constant proportions $r_r$ and $e_r$ of the
total number of recombination and error variables, respectively.
The base values of the maximum recombination rate $r_r$ and the maximum
error rate $e_r$ is $0.012$.
Since a random generator can produce instances where the actual error or
recombination rate is larger than $\theta$ and $\eps$, the actual values of
$r$ and $e$ given to \progr must be chosen appropriately.

Table~\ref{tab:rnd} summarizes the accuracy and the performance of
\progr on two series that are representative of this experimental part.
Due to space constraints, we presented the complete table in the
supplementary material (Tab.~A.1) and we only sketch the main
conclusions.
We notice that the average genotype imputation error rate is
always below
21\%, the average haplotyping error rate is always below 8\% and the average
running time is always below 9 minutes, while the instance that took the
longest time terminated in 52 minutes.
The best case for each row is boldfaced in order to show the effects that
parameter variations has on the outcome. The main noteworthy, albeit
predictable, observation is that larger pedigrees require more computational
resources, but result in a more accurate prediction.

\begin{table}[tp!]

\caption{Summary of accuracy and performance obtained by \progr on
  randomly generated instances.
  Each table refers to a series of tests where only one parameter has
  been varied.
  The base values of the parameters are:
$n=50$,
$m=50$,
$\theta=0.005$,
$\eps=0.005$,
$\mu=0.05$. The best result for each row is boldfaced.}
\label{tab:rnd}

\centering\scriptsize

\subtable[Increasing pedigree size ($n$)]{
\begin{tabularx}{0.95\linewidth}{l W W W}
  \toprule
  Pedigree size $n=$ & 50 & 100 & 200 \\
  \midrule
  No. of completed instances & 100/100 & 100/100 & 100/100 \\
  Avg. genotype imputation error rate & 0.161 & 0.151 & \textbf{0.149} \\
  Avg. haplotyping error rate & 0.043 & 0.039 & \textbf{0.033} \\
  Avg. haplotyping error rate (wo/missing) & 0.037 & 0.034 & \textbf{0.027} \\
  Avg. running time (in seconds) & \textbf{15.1} & 82.3 & 359.4 \\
  Max. running time (in seconds) & \textbf{72.3} & 479.9 & 1129.3 \\
  \bottomrule
\end{tabularx}
\label{tab:rnd:size}
}

\subtable[Increasing genotype length $(m)$]{
\begin{tabularx}{0.95\linewidth}{l W W W}
  \toprule
  Genotype length $m=$ & 50 & 100 & 200 \\
  \midrule
  No. of completed instances & 100/100 & 100/100 & 100/100 \\
  Avg. genotype imputation error rate & \textbf{0.170} & 0.174 & 0.174 \\
  Avg. haplotyping error rate & \textbf{0.043} & 0.049 & 0.063 \\
  Avg. haplotyping error rate (wo/missing) & \textbf{0.038} & 0.045 & 0.059 \\
  Avg. running time (in seconds) & \textbf{17.0} & 93.6 & 505.0 \\
  Max. running time (in seconds) & \textbf{85.7} & 770.4 & 3127.5 \\
  \bottomrule
\end{tabularx}
\label{tab:rnd:length}
}
\end{table}


\subsubsection*{Comparison with a State-of-the-Art Method.}
\label{sec:experim:cmp}

In this second part, we compared accuracy and performance of \progr with
another state-of-the-art approach for the
haplotype inference problem on pedigrees: MePhase~\cite{mephase}.
A second approach, PedPhase 3.0~\cite{LiLi09}, was initially considered.
A preliminary test revealed that PedPhase
terminated without giving a solution or giving an error message on a
significant subset of instances and we preferred to not include it in
this comparison.
MePhase is a heuristic algorithm for the \emph{haplotype configuration
with mutation and errors} problem on \emph{tree} pedigrees (\ie
pedigrees such that each pair of individuals are connected by at most
one directed path) and it is based on an ILP formulation derived
from a system of linear equations over $\Z_2$~\cite{XiaoLXJ09}.
This formulation of the HI problem asks for a haplotype configuration of
a given genotyped pedigree allowing mutations, genotyping errors, and
missing genotypes, while in this work we allow recombinations,
genotyping errors, and missing genotypes.
As a consequence, the comparison with MePhase involved randomly
generated haplotype configurations on tree pedigrees (opposed to the
general pedigrees used in the first part) \emph{without} mutations and
recombinations.
To ensure a proper comparison even if our approach and MePhase were
designed for slightly different problems, the experimental evaluation is
similar to the one described by the authors of MePhase in~\cite{mephase}.
Tree pedigrees have been generated using the random generator proposed by
Thomas and Cannings~\cite{treeped-gen}.
Random haplotype configurations have been assigned to the tree pedigrees just
as in the previous experimental part.
In this case, we analyzed the effects of the following problem parameters
on the accuracy and performance of MePhase and \progr:
\emph{average nuclear family size} (denoted with $f$, and represents the
average number of individuals that compose a nuclear family),
\emph{genotype length} ($m$), \emph{error probability} ($\eps$), and
\emph{missing probability} ($\mu$).
We generated five different tree pedigrees for each value of $f$ ranging from
$3$ to $6$.
For each pedigree and for each combination of the
remaining parameters $(m, \eps, \mu)$, six random haplotype
configurations have been computed.
We considered the following set of parameter values: $m \in \{ 50,
100\}$, $\eps \in \{ 0, 0.005, 0.01 \}$, and $\mu \in \{ 0, 0.05,
0.1, 0.2 \}$ for a total of $2880$ instances.
Table~\ref{tab:cmp} summarizes the results of the comparison.
Accuracy of the two approaches in imputing the missing genotypes and
reconstructing the haplotype configuration is similar:
\progr is definitely more accurate than MePhase on smaller families
($f\ge 4$), while MePhase is slightly more accurate
on larger families ($f\ge 5$).
A likely reason is  that MePhase  computes
haplotype configurations with fewer genotyping errors than \progr on
larger families, as witnessed by the fact that, on those families, the average
number of errors is smaller for MePhase than for \progr. Also, on those
families, precision and recall (\ie the fraction of original  and computed
errors that have been correctly identified) are better for MePhase than
for \progr.

However, \progr is considerably more efficient than MePhase:
the difference is most remarkable on large families ($f=6$) where
MePhase took $1092$ seconds on average compared to the $23$ seconds
required by \progr.
Moreover, MePhase was not able to solve $266$ of the original $2880$
instances ($9.2\%$) within a time limit of an hour for each instance,
while \progr completed the whole dataset within the same time limit.
MePhase could not solve within the time limit $31.2\%$ of the instances with $f=6$.
Overall, \progr is much faster and more scalable than MePhase while
maintaining comparable accuracy.

\begin{table}[t!]
  \scriptsize\centering

  \caption{Results of the experimental comparison between
    MePhase~\cite{mephase} and \progr.
    The first column ($f$) indicates the average size of nuclear family,
    the second reports the average number of errors present in the
    generated haplotype configuration (original errors),
    the next four columns (and the last one) are defined as the rows of
    Table~\ref{tab:rnd}.
    The 7th column indicates the average number of errors present in
    the reconstructed haplotype configuration (computed errors).
    \emph{Precision} and \emph{recall} are defined as the proportion of
    original (computed, resp.) errors that have been correctly
    identified.
    Since MePhase was not able to solve all the instances within the
    1-hour time limit, we also computed the measures obtained by \progr
    restricted to the subset of instances completed by MePhase.
  }
  \label{tab:cmp}

  \begin{tabularx}{0.98\textwidth}{T T W W W W W W W W W W W W W W W W}
\toprule
        &       & \multicolumn{8}{c}{MePhase} \\
\cmidrule{3-10}
        &       &       & \multicolumn{1}{c}{genot.}    &       & \multicolumn{1}{c}{haplot.}   &       &       &       & \multicolumn{1}{c}{avg.}      \\
        &       &       & \multicolumn{1}{c}{imput.}    & \multicolumn{1}{c}{haplot.}   & \multicolumn{1}{c}{error}     & \multicolumn{1}{c}{no.~of}    &       &       & \multicolumn{1}{c}{running}   \\
        & \multicolumn{1}{c}{no.~of}    & \multicolumn{1}{c}{compl.}    & \multicolumn{1}{c}{error}     & \multicolumn{1}{c}{error}     & \multicolumn{1}{c}{rate}      & \multicolumn{1}{c}{comp.}     &       &       & \multicolumn{1}{c}{time}      \\
$f$     & \multicolumn{1}{c}{errors}    & \multicolumn{1}{c}{instan.}   & \multicolumn{1}{c}{rate}      & \multicolumn{1}{c}{rate}      & \multicolumn{1}{c}{wo/miss}   & \multicolumn{1}{c}{errors}    & \multicolumn{1}{c}{precision} & \multicolumn{1}{c}{recall}    & \multicolumn{1}{c}{(sec)}     \\
\otoprule
3       & 17.1  & 719   & 0.234 & 0.044 & 0.024 & 21.1  & 0.503 & 0.376 & 5.8   \\
4       & 17.1  & 713   & 0.092 & 0.017 & 0.007 & 17.1  & 0.751 & 0.627 & 16.0  \\
5       & 16.1  & 687   & 0.048 & 0.008 & 0.003 & 15.8  & 0.854 & 0.788 & 83.1  \\
6       & 13.8  & 495   & 0.022 & 0.003 & 0.001 & 13.6  & 0.926 & 0.921 & 1092.1        \\
\midrule
Overall & 16.2  & 2614  & 0.106 & 0.019 & 0.009 & 17.2  & 0.739 & 0.630 & 234.6 \\
\midrule
        &       & \multicolumn{8}{c}{\progr} \\
\cmidrule(lr){1-2}\cmidrule(l){3-10}
3       & 17.1  & 720   & 0.123 & 0.025 & 0.014 & 16.8  & 0.663 & 0.643 & 15.2 \\
4       & 17.1  & 720   & 0.072 & 0.016 & 0.009 & 21.0  & 0.654 & 0.745 & 19.5 \\
5       & 16.1  & 720   & 0.049 & 0.011 & 0.007 & 21.6  & 0.629 & 0.765 & 19.7 \\
6       & 13.8  & 720   & 0.024 & 0.006 & 0.004 & 22.0  & 0.678 & 0.841 & 22.6 \\
\midrule
Overall   & 16.2  & 2880  & 0.067 & 0.015 & 0.008 & 20.3  & 0.657 & 0.737 & 19.2 \\
\midrule
        &       & \multicolumn{8}{c}{\progr (on instances completed also by MePhase)} \\
\cmidrule(lr){1-2}\cmidrule(l){3-10}
3       & 17.1  & 719   & 0.123 & 0.025 & 0.014 & 16.8  & 0.663 & 0.643 & 15.2 \\
4       & 17.1  & 713   & 0.072 & 0.016 & 0.009 & 20.9  & 0.654 & 0.744 & 19.3 \\
5       & 16.1  & 687   & 0.050 & 0.011 & 0.007 & 20.2  & 0.627 & 0.762 & 16.5 \\
6       & 13.8  & 495   & 0.024 & 0.006 & 0.004 & 17.8  & 0.671 & 0.839 & 14.8 \\
\midrule
Overall   & 16.2  & 2614  & 0.071 & 0.015 & 0.009 & 19.0  & 0.653 & 0.727 & 16.6 \\
\bottomrule

  \end{tabularx}

\end{table}


\subsubsection*{Evaluation on a Real Genotyped Pedigree.}
\label{sec:experim:real}

In the last part, we evaluated our approach on a real genotyped
pedigree.
The pedigree describes part of a dairy cattle population that has been
obtained from the Italian Brown Swiss Breeders Association (ANARB).
Genotypes have been obtained from a BovineSNP50 BeadChip and were
restricted to 2570 loci on Chromosome 6.
The pedigree (represented in the Supplementary Figure~S.1)
contains $207$ individuals -- $130$ males and $77$
females -- with $93$  founders.
Unlike humans pedigrees, livestock pedigrees are composed by large
families of half-siblings.
In fact, the pedigree contains
two bulls that generated $17$ and $15$ offspring, respectively, while
other $12$ bulls generated $50$ offspring.
A second typical characteristic  of livestock pedigrees is the presence of
several individuals that are not genotyped.
The reason is twofold:
the introduction of genotyping
technologies for livestock species is recent.
Therefore, most of the animals of the pedigree are not alive and cannot
be genotyped.
Moreover,
the extraction of genotypes is quite expensive, hence it is performed only on
animals of higher commercial value, such as breeding animals.
In our pedigree we have the genotypes of only $105$ males (and no females).
Consequently almost half of all genotypes are missing.

The aim of this part of the experimentation is twofold:
(i)  to validate the minimum-recombinant and
minimum-error model implicitly assumed in the formulation of the
\reHCstar problem and, (ii) to empirically
prove that \progr can handle large, real-world pedigrees with a lot of non-genotyped individuals.

We prepared $6$ instances from the original data by selecting $6$
subsets of $50$ original markers spaced at different distances $d$.
The first subset has distance $d=1$ or, in other words, is composed by
the first $50$ original markers.
The second subset has distance $d=2$ (\ie we selected every other marker),
the third has distance $d=10$ (\ie we selected one marker every ten
original subsequent markers), and the other three have distance $d=15,
20, 25$, respectively.
The genotypes used in the experimentation have been obtained by
restricting the original genotypes to the subsets of selected markers.
This process simulates genotypes collected at different ``densities''.
Since the closer the loci, the stronger their linkage, we
expect that high-density genotypes (the ones with distance
$d=1$ and $d=2$) can be solved with only a few recombinations and/or errors,
while low-density genotypes should require a
larger amount of recombinations and/or errors.
To test this hypothesis, we ran \progr over the $6$ instances with
different maximum recombination and error rates in order to find the
haplotype configuration with the smallest number of recombinations and
errors.

The results of this experiment (Table~\ref{tab:real}) reveal a clear
trend:
genotypes with higher intra-marker distance (\ie lower densities)
require considerably more recombinations and/or errors than high-density
genotypes.
In fact, it was possible to find a haplotype configuration with a single
genotyping error (and no recombination) that solves the instance with the highest density
($d=1$), while $45$ errors (or $50$ recombinations) were needed in the
solution of the medium-density instance ($d=15$), and $89$ errors (or
$6$ recombinations and $81$ errors) were needed in the lowest-density
instance ($d=25$).
Lower-density instances ($d\ge 15$) have not been solved within a time limit of $3$
hours for some particular choices of maximum recombination and error
rates.

\begin{table}[t!]
  \scriptsize\centering
  \caption{Summary of the minimum number of recombinations and errors
    needed to solve the livestock pedigree at 6 different marker distances
    $d=1, 2, 10, 15, 20, 25$.
    The first two columns indicate the maximum recombinations
    and  error rates given in input to \progr.
    Each other cell has either indicated the number of recombinations
    ($r$) and errors ($e$) in the resulting haplotype configuration or
    ``no solution'' if no solution exists within the given recombination
    and error rates or ``timeout'' if no solution has been found within the
    time limit.
    Blank cells indicate instances already solved with a more stringent
    choice of recombination and error rates.
  }
  \label{tab:real}
  \begin{tabularx}{0.98\textwidth}{T T T T T T T T}
    \toprule
    Recomb. & Error & \multicolumn{6}{c}{Marker distances $d$} \\
    \cmidrule(lr){3-8}
    rate $r_r$ & rate $e_r$ & 1  & 2 & 10 & 15 & 20 & 25 \\
\otoprule
0.0 & 0.0 & no sol. & no sol. & no sol. & no sol. & no sol. & no sol. \\
\midrule
0.0005  & 0.0   & $r=4$ & $r$=5 & no sol.       & no sol.       & no sol.       & no sol.\\
0.001   & 0.0   &       &       & no sol.       & no sol.       & no sol.       & no sol.\\
0.005   & 0.0   &       &       & $r$=25        & $r$=50        & $r$=52        & $r$=53\\
\midrule
0.0     & 0.0005        & $e$=1 & no sol.       & no sol.       & no sol.       & no sol.       & no sol.\\
0.0     & 0.001 &       & $e$=5 & no sol.       & no sol.       & no sol.       & no sol.\\
0.0     & 0.005 &       &       & $e$=27        & timeout       & no sol.       & timeout\\
0.0     & 0.01  &       &       &       & $e$=45        & timeout       & timeout\\
0.0     & 0.02  &       &       &       &       & $e$=94        & $e$=89\\
\midrule
0.0005  & 0.0005        & $r$=0, $e$=1  & $r$=5, $e$=0  & no sol.       & no sol.       & no sol.       & no sol.\\
0.0005  & 0.005 &       &       & $r$=6, $e$=20 & timeout       & timeout       & timeout\\
0.0005  & 0.01  &       &       &       & $r$=0, $e$=45 & timeout       & timeout\\
0.0005  & 0.02  &       &       &       &       & $r$=0, $e$=94 & $r$=6, $e$=81\\
\midrule
0.001   & 0.0005        &       &       & $r$=12, $e$=3 & no sol.       & no sol.       & no sol. \\
0.001   & 0.001 &       &       & $r$=11, $e$=6 &  timeout      & timeout       &  no sol.      \\
\midrule
0.005   & 0.0005        &       &       &       & $r$=40, $e$=3 & $r$=48, $e$=3 & $r$=50, $e$=1\\
\bottomrule
  \end{tabularx}
\end{table}


Our overall conclusion is that
the results of the three experimental
parts support the soundness of the model and the feasibility of the
approach.
In fact, \progr has computed solutions with good accuracy on
simulated instances even in limit cases (pedigrees with many
recombinations and errors and/or many untyped loci) and has found
haplotype configurations with a few recombinations and
errors on some real instances.
The comparison with MePhase,
revealed that \progr has a comparable accuracy, but it is much faster and
better scales to large and complex pedigrees.

\subsubsection*{Acknowledgments.}
We would like to thank Wei-Bung Wang for valuable discussions and
sharing the MePhase code with us, and Dr.~Enrico Santus, director of
``Associazione Nazionale Allevatori Razza Bruna'', Bussolengo, Italy,
for providing the pedigree and the genotypes used in this work.

\clearpage
\appendix
\section*{Supplementary Material}

\textbf{Supplementary Table~S.1.}
  Summary of accuracy and performance obtained by \progr on
  randomly generated instances.
  Each table refers to a series of tests where only one parameter has
  been varied.
  The base values of the parameters are:
  pedigree size
$n=50$,
  genotype length
$m=50$,
  recombination probability
$\theta=0.005$,
  error probability
$\eps=0.005$,
  and missing probability
$\mu=0.05$. The best result for each row is boldfaced.

\begin{center}
\scriptsize

{\small (a) Increasing pedigree size ($n$)}\\[0.6ex]
\begin{tabularx}{0.95\linewidth}{l W W W}
  \toprule
  Pedigree size $n=$ & 50 & 100 & 200 \\
  \midrule
  No. of completed instances & 100/100 & 100/100 & 100/100 \\
  Avg. genotype imputation error rate & 0.161 & 0.151 & \textbf{0.149} \\
  Avg. haplotyping error rate & 0.043 & 0.039 & \textbf{0.033} \\
  Avg. haplotyping error rate (wo/missing) & 0.037 & 0.034 & \textbf{0.027} \\
  Avg. running time (in seconds) & \textbf{15.1} & 82.3 & 359.4 \\
  Max. running time (in seconds) & \textbf{72.3} & 479.9 & 1129.3 \\
  \bottomrule
\end{tabularx}

\medskip
{\small (b) Increasing genotype length $(m)$}\\[0.6ex]
\begin{tabularx}{0.95\linewidth}{l W W W}
  \toprule
  Genotype length $m=$ & 50 & 100 & 200 \\
  \midrule
  No. of completed instances & 100/100 & 100/100 & 100/100 \\
  Avg. genotype imputation error rate & \textbf{0.170} & 0.174 & 0.174 \\
  Avg. haplotyping error rate & \textbf{0.043} & 0.049 & 0.063 \\
  Avg. haplotyping error rate (wo/missing) & \textbf{0.038} & 0.045 & 0.059 \\
  Avg. running time (in seconds) & \textbf{17.0} & 93.6 & 505.0 \\
  Max. running time (in seconds) & \textbf{85.7} & 770.4 & 3127.5 \\
  \bottomrule
\end{tabularx}

\medskip
{\small (c) Increasing recombination probability $(\theta)$}\\[0.6ex]
\begin{tabularx}{0.95\linewidth}{l W W W W}
  \toprule
  Recombination probability $\theta=$ & 0.0 & 0.005 & 0.01 & 0.02 \\
  \midrule
  No. of completed instances & 100/100 & 100/100 & 100/100 & 100/100 \\
  Avg. genotype imputation error rate & 0.174 & \textbf{0.173} & 0.174 & 0.185 \\
  Avg. haplotyping error rate & \textbf{0.047} & 0.050 & 0.057 & 0.077 \\
  Avg. haplotyping error rate (wo/missing) & \textbf{0.040} & 0.047 & 0.052 & 0.072 \\
  Avg. running time (in seconds) & \textbf{5.0} & 7.0 & 13.0 & 10.5 \\
  Max. running time (in seconds) & \textbf{14.7} & 50.1 & 75.7 & 62.6 \\
  \bottomrule
\end{tabularx}

\medskip
{\small (d) Increasing error probability $(\eps)$}\\[0.6ex]
\begin{tabularx}{0.95\linewidth}{l W W W}
  \toprule
  Error probability $\eps=$ & 0.0 & 0.005 & 0.01 \\
  \midrule
  No. of completed instances & 100/100 & 100/100 & 100/100 \\
  Avg. genotype imputation error rate & 0.172 & \textbf{0.166} & \textbf{0.166} \\
  Avg. haplotyping error rate & 0.048 & \textbf{0.047} & 0.048 \\
  Avg. haplotyping error rate (wo/missing) & \textbf{0.042} & \textbf{0.042} & 0.044 \\
  Avg. running time (in seconds) & \textbf{12.8} & 28.9 & 74.2 \\
  Max. running time (in seconds) & \textbf{55.5} & 109.8 & 485.2 \\
  \bottomrule
\end{tabularx}

\medskip
{\small (e) Increasing missing probability $(\mu)$}\\[0.6ex]
\begin{tabularx}{0.95\linewidth}{l W W W W}
  \toprule
  Missing probability $\mu=$ & 0.0 & 0.05 & 0.1 & 0.2 \\
  \midrule
  No. of completed instances & 100/100 & 100/100 & 100/100 & 100/100 \\
  Avg. genotype imputation error rate & ---\ \parbox{0pt}{} & \textbf{0.167} & 0.183 & 0.207 \\
  Avg. haplotyping error rate & \textbf{0.032} & 0.041 & 0.052 & 0.071 \\
  Avg. haplotyping error rate (wo/missing) & \textbf{0.032} & 0.034 & 0.041 & 0.052 \\
  Avg. running time (in seconds) & 16.6 & 15.3 & 17.9 & \textbf{14.0} \\
  Max. running time (in seconds) & 79.7 & 76.0 & 89.0 & \textbf{69.2} \\
  \bottomrule
\end{tabularx}
\end{center}


\renewcommand{\figurename}{Supplementary Figure S.\!}
\begin{sidewaysfigure}\centering
  \includegraphics[width=\textheight]{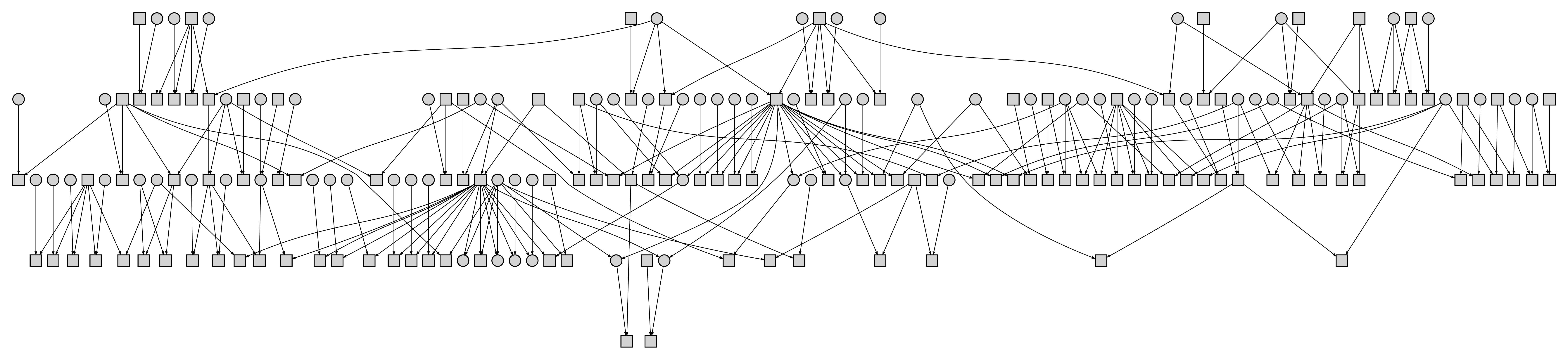}
  \caption{Pedigree graph of the real dairy cattle population used in
    the experimental evaluation of \progr.
    Male individuals are conventionally represented as boxes, while
    females as circles. Direct edges connect an individual with its
    child.}
\end{sidewaysfigure}

\end{document}